\documentclass{article}
\usepackage[T1]{fontenc}
\usepackage{spconf}       
\usepackage{amsmath}      
\usepackage{amssymb}      
\usepackage{graphicx}     
\usepackage{algorithm}    
\usepackage{algpseudocode}
\usepackage{hyperref}     
\usepackage{xspace}       
\usepackage{caption}      
\usepackage{subcaption}   
\usepackage{comment}      
\usepackage{multirow}     
\usepackage{siunitx}      
\usepackage{makecell}     
\usepackage{array}        
\usepackage{threeparttable}  
\usepackage{booktabs}     

\title{Analytic Incremental Learning for Sound Source Localization with Imbalance Rectification}
\name{Zexia Fan${^1}$, Yu Chen${^{1,2}}$, Qiquan Zhang${^3}$, Kainan Chen${^4}$, Xinyuan Qian${^1}{^*}$
\thanks{This work is supported by National Natural Science Foundation of China (62306029), Beijing Natural Science Foundation (L233032) and Young Elite Scientists Sponsorship Program of the Beijing High Innovation Plan.\\ \hspace*{1.5em}*Corresponding author}}
\address{${^1}$University of Science and Technology Beijing, Beijing, China
\\
${^2}$The Chinese University of Hong Kong, Shenzhen, China
\\
${^3}$Tongyi Speech Lab, Alibaba, China
\\
${^4}$Eigenspace GmbH, Germany
}
\begin{document}
\ninept

\maketitle
\begin{abstract}
\textcolor{black}{Sound source localization (SSL) demonstrates remarkable results in controlled settings but struggles in real-world deployment due to dual imbalance challenges: intra-task imbalance arising from long-tailed direction-of-arrival (DoA) distributions, and inter-task imbalance induced by cross-task skews and overlaps.
 These often lead to \textit{catastrophic forgetting}, significantly degrading the localization accuracy.
To \textcolor{black}{mitigate} these issues, we propose a unified framework with two key innovations. Specifically we design a GCC-PHAT-based data augmentation (GDA) method that leverages peak characteristics to alleviate intra-task distribution skews. We also propose an Analytic dynamic imbalance rectifier (ADIR) with task-adaption regularization, which enables analytic updates that adapt to inter-task dynamic. On the SSLR benchmark, our proposal achieves state-of-the-art (SoTA) results of 89.0\% accuracy, 5.3$^{\circ}$ mean absolute error, and 1.6 backward transfer, demonstrating robustness to evolving imbalances without exemplar storage. {Code and data will be released.}}

\end{abstract}
\vspace{-1.5mm}
\vspace{-4mm}
\section{Introduction}
\vspace{-3mm}
\label{sec:intro}


Sound Source Localization (SSL) aims to estimate the DoA of sound sources from multichannel audio. In the context of human-robot interaction, SSL is fundamental for selective auditory attention~\cite{chen2024locselect}, automatic speech recognition \cite{conmamba}, speech enhancement~\cite{deepmmse,tfase,11030293}, and speaker extraction~\cite{xu2020speaker} in multi-speaker and noisy scenarios. Traditional signal processing approaches rely on idealized acoustic assumptions and derive closed-form solutions, such as GCC-PHAT~\cite{knapp1976generalized}, MUSIC~\cite{MUSIC}, and SRP-PHAT~\cite{SRPPHAT}. They often perform well under controlled scenarios, but struggle with challenging acoustic conditions such as noisy and strong reverberant environments. Deep learning-based supervised neural SSL has recently demonstrated significant advancements. Existing approaches can be broadly grouped into hybrid~\cite{he2018deep,music1,srp2,perotin2019crnn} and pure neural solutions~\cite{chakrabarty2017multi,qian2021doa,cnnlstm_doa,seldnet}. The former leverages a deep neural network (DNN) to enhance specific modules of traditional pipelines. While the latter optimizes DNNs to estimate DoA in an end-to-end fashion.

{In practice, labeled SSL data is scarce and arrive incrementally, as in robotic audition where DoA annotations evolve over time. A critical problem in this incremental setting is catastrophic forgetting, where the model loses previously acquired knowledge when adapting to new tasks. Class-incremental learning (CIL) has been explored to address this issue, which seeks to integrate new classes while preserving old knowledge. Existing CIL generally spans three paradigms: regularization-based approaches~\cite{lwf,schwarz2018progress} that constrain weight updates to retain information, replay-based approaches~\cite{icarl,cermelli2020modeling} that store exemplars for knowledge distillation, and analytic approaches~\cite{zhuang2023gkeal,zhuang2024dsal,gacl,zhuang2024foal} that exploit closed-form solutions for computational efficiency. Building on these foundations, SSL-CIL~\cite{qian_cil} introduced the first CIL framework for SSL, yet CIL method remain challenged by complex real-world conditions such as repeated class occurrences and class imbalance problems.}

These limitations become particularly acute in the generalized CIL (GCIL) setting for SSL, which faces a fundamental challenge: the long-tailed class distributions of DoA data. This manifests in two forms: 1) \textit{Intra-task imbalance} arises within each task, where certain directions dominate training data while minority directions remain severely underrepresented~\cite{liu2022longtail,air}; 2) \textit{Inter-task imbalance} occurs across tasks, with uneven or overlapping class distributions, which induces dynamic biases and exacerbates forgetting~\cite{raja2024exploring}. Together, these imbalances make GCIL for SSL particularly difficult, especially when privacy constraints prohibit storing past data.


To tackle these challenges, we propose a novel unified framework, namely SSL-GCIL, that addresses both intra-task and inter-task imbalances. For intra-task imbalance, we introduce a GDA method that exploits GCC-PHAT peak statistics to synthesize samples for tail classes, preserving inter-microphone correlations without requiring external data. For inter-task imbalance, we design ADIR, an analytic classifier whose core adaptive regularization is guided by task-specific statistics (e.g., the Gini coefficient) to effectively mitigate cross-task skews and forgetting~\cite{air}. 
Together, these components form a lightweight, adaptive SSL framework tailored for GCIL.
In summary, our contributions are summarized as:
\begin{itemize}
    \vspace{-1.5mm}
    \item {Our proposal is the first SSL framework} that address both intra-task and inter-task imbalances in the GCIL setting, providing an effective solution for long-tailed distributions which alleviate catastrophic forgetting.
    \vspace{-1.5mm}
    \item We introduce GDA that exploits GCC-PHAT peak statistics to synthesize tail-class samples, alleviating intra-task imbalance while maintaining statistical consistency and privacy.
    \vspace{-1.5mm}
    \item We propose an ADIR, a unified analytic framework that dynamically responds to inter-task class overlaps and distribution skews through its task-adaptive regularization. 
\end{itemize}
\vspace{-1.5mm}
We conduct comprehensive experiments on {the  SSLR benchmark, the experimental results demonstrate the superiority of our proposed SSL-GCIL over SoTA baselines, achieving improvements of 3.2\% in accuracy, 1.2$^\circ$ in mean absolute error, and 2.5 in backward transfer.}

\begin{figure*}[t]
\centering
\includegraphics[width=0.92\textwidth]{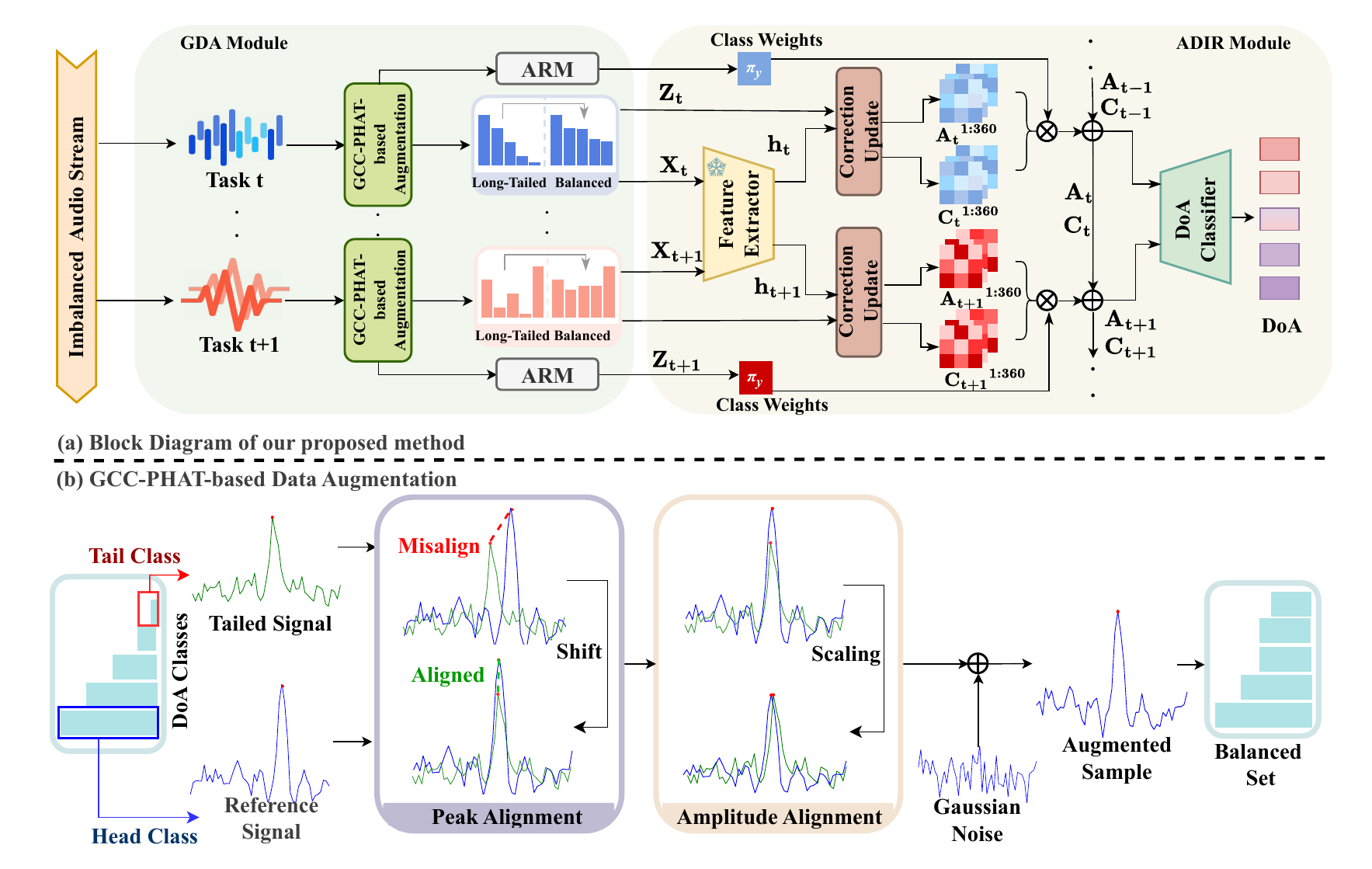}
\caption{(a) Block Diagram of our proposed method and  (b) GCC-PHAT-based Data Augmentation ($\bigoplus$ indicates addition). }
\label{fig:framework}
\vspace{-1.5em}
\end{figure*}
\vspace{-4mm}
\section{Method}
\label{sec:method}


In this section, we first formalize the problem of GCIL for SSL and then introduce our proposed method. As illustrated in Fig.~\ref{fig:framework}(a), the framework is composed of two main components: a GDA module for feature enrichment and an ADIR module with a frozen MLP feature extractor to map GCC-PHAT features to DoA probability.

\vspace{-3mm}
\subsection{Problem Formulation}
\label{ssec:problem}

We formulate SSL as a GCIL problem, where data arrives in tasks, each containing new and old sound direction classes with imbalanced distributions. Let $\mathcal{D}_t = \{(\mathbf{x}_i^{(t)}, \mathbf{y}_i^{(t)}, \mathbf{z}_i^{(t)})\}_{i=1}^{N_t}$ denote the training set of task $t \in \{1, \dots, T\}$, where $\mathbf{x}_i^{(t)} \in \mathbb{R}^{P \cdot D_a}$ is the GCC-PHAT feature derived from a $P$-pair microphone array,  $D_a$ being the feature dimension per pair, $\mathbf{y}_i^{(t)} \in \{0,1\}^{360}$ is the one-hot label indicating the true DoA in $[0^\circ, 360^\circ)$, and $\mathbf{z}_i^{(t)} \in \mathbb{R}^{360}$ is the Gaussian-distributed label generated from $\mathbf{y}_i^{(t)}$ with a peak at the true DoA and standard deviation $\sigma=5^\circ$. The classes in $\mathcal{D}_t$ follow a long-tailed distribution with up to 60 classes per task. The goal is to train a model $f_\theta: \mathbb{R}^{P \cdot D_a} \to \mathbb{R}^{360}$ minimizing cumulative loss over tasks without revisiting prior data:
\vspace{-3mm}

\begin{equation}
\mathcal{L} = \sum_{t=1}^T \mathbb{E}_{(\mathbf{x}, \mathbf{z}) \sim \mathcal{D}_t} \left[ \ell(f_\theta(\mathbf{x}), \mathbf{z}) \right],
\end{equation}
where $\ell$ is binary cross-entropy loss, addressing catastrophic forgetting and intra-task imbalance without exemplar storage.
\vspace{-3mm}
\subsection{GCC-PHAT-based Data Augmentation}
\label{ssec:augmentation}

The GCC-PHAT estimates source directions from inter-microphone time delays. For signals with Fourier transforms $X(f)$ and $Y(f)$, it is defined as
\vspace{-1.5mm}
\begin{equation}
R_{xy}(\tau) = \int_{-\infty}^{\infty} \frac{X(f) Y^*(f)}{|X(f) Y^*(f)|} e^{j 2\pi f \tau} \, df,
\end{equation}
where $\tau$ denotes the lag. Empirically, the peaks of GCC-PHAT dominate SSL performance, as they encode critical time-difference cues \cite{qian2021doa}. To mitigate intra-task imbalance under long-tailed distributions \cite{acil}, we design a domain-specific augmentation that manipulates these peaks while preserving their statistical structure.

For each task $t$, the feature vector $\mathbf{x}_i^{(t)} \in \mathbb{R}^{P \cdot D_a}$ is partitioned into $P$ segments (one per microphone pair), each of dimension $D_a$. Let $N_c^{(t)}$ denote the number of samples in class $c$ at task $t$. For class $c$, we compute mean and variance of peak positions $\{p_{c,k}\}$ and amplitudes $\{a_{c,k}\}$ across segments. Let $M_t = \max_c N_c^{(t)}$ be the largest class size and $\alpha=0.5$ the augmentation rate. For any class with $N_c^{(t)} < \alpha M_t$, we generate $
K_c = \lceil \alpha M_t - N_c^{(t)} \rceil$
new samples using Algorithm~\ref{alg:augmentation}. The procedure shifts and rescales peaks of abundant-class features to match the statistics of tail classes, while injecting low-level noise for diversity. This produces $\tilde{\mathcal{D}}_t$, balancing tail representation without distorting inter-microphone correlations.

\begin{algorithm}[t]
\caption{GDA for Tail Class $c$}
\label{alg:augmentation}
\begin{algorithmic}[1]
\Require Base feature $\mathbf{x}_b \in \mathbb{R}^{P \cdot D_a}$ from abundant class $c'$, statistics $\{p_{c,k}, a_{c,k}\}_{k=1}^P$
\Ensure Augmented sample $\mathbf{x}_{\text{new}}$
\State Initialize $\mathbf{x}_{\text{new}} \leftarrow \mathbf{0}^{P \cdot D_a}$
\For{segment $k = 1,\dots,P$}
    \State $\Delta p_k \leftarrow p_{c,k} - p_{c',k}$
    \State Extract $\mathbf{x}_b^{(k)} \in \mathbb{R}^{D_a}$ from $\mathbf{x}_b$
    \State Cyclically shift $\mathbf{x}_b^{(k)}$ by $\Delta p_k$
    \State Scale amplitudes to match $a_{c,k}$
    \State Add Gaussian noise $\mathcal{N}(0, \sigma_n)$, with $\sigma_n = 0.05 \cdot \max(\mathbf{x}_b^{(k)})$
    \State Assign segment to $\mathbf{x}_{\text{new}}$
\EndFor\\
\Return $\mathbf{x}_{\text{new}}$
\end{algorithmic}
\end{algorithm}


\vspace{-3mm}
\subsection{Network Architecture}
\label{ssec:architecture}

We adopt a network: a frozen feature extractor followed by a task-adaptive classifier. The input $\mathbf{x} \in \mathbb{R}^{P \cdot D_a}$ is mapped to logits $\hat{\mathbf{z}} \in \mathbb{R}^{360}$ via
\begin{equation}
    \mathbf{h} = F_{\text{MLP}}(\mathbf{x}; \mathbf{W}_{\text{MLP}}), \quad
    \hat{\mathbf{z}} = F_{\text{cls}}(\mathbf{h}; \mathbf{W}_{\text{t}}).
\end{equation}
{where $F_{\text{MLP}}$ is a three-layer MLP, with each layer consisting of batch normalization, ReLU and dropout. $F_{\text{cls}}$ is a fully-connected layer with dropout.} The feature extractor $\mathbf{W}_{\text{MLP}}$ is only trained in task 1. For subsequent tasks, $\mathbf{W}_{\text{MLP}}$ is frozen and only $\mathbf{W}_{\text{t}}$ is updated.

\vspace{-3mm}
\subsection{Integration of Analytic Dynamic Imbalance Rectifier}
\label{ssec:air}

We design the ADIR \cite{air} for the GCIL setting. Unlike conventional CIL where classes across tasks are disjoint, GCIL allows classes to reappear, requiring per-class storage of auto-correlation and cross-correlation matrices. {Notably, our framework operates without storing any raw audio data or exemplars, ensuring compliance with privacy constraints in real-world applications.} 

The detailed procedure is shown in Algorithm~\ref{alg:air}. For each task $t$, given the frozen MLP features $\mathbf{H}_t \in \mathbb{R}^{N_t \times 1000}$ and Gaussian-smoothed labels $\mathbf{Z}_t$, the Adaptive Re-weighting Module (ARM)\cite{air} estimates the optimal weight matrix through
\begin{equation}
\mathbf{{W}}^*_t = \arg\min_{\mathbf{W}_t} \sum_{c=1}^{360} \pi_c \|\mathbf{Z}_t^{(c)} - \mathbf{H}_t^{(c)} \mathbf{W}_t\|_F^2 + \gamma_t \|\mathbf{W}_t\|_F^2,
\end{equation}
where $\pi_c = 1/N_c$ is the re-weighting factor for class c, and $\gamma_t$ is an adaptive regularization parameter to enhance stability. Specifically, 
    $\gamma_t = \gamma_0 \cdot e^{\alpha \cdot (\text{Gini}_t - 0.5)}$,
with $\gamma_0=100$, $\alpha=2$, and $\text{Gini}_t$ is the Gini coefficient computed as (for $n=60$ classes):
\begin{equation}
    \text{Gini}_t = \frac{\sum_{i=1}^n \sum_{j=1}^n |p_i - p_j|}{2n \sum_{i=1}^n p_i}.
    \label{eq:gini}
\end{equation}
the closed-form update is given by
\begin{equation}
\mathbf{{W}}^*_t = \left( \sum_{c} \pi_c \mathbf{A}^{(c)} + \gamma_t \mathbf{I} \right)^{-1} \left( \sum_{c} \pi_c \mathbf{C}^{(c)} \right),
\end{equation}
with $\mathbf{A}^{(c)} = \mathbf{H}_t^{(c)\top}\mathbf{H}_t^{(c)}$ and $\mathbf{C}^{(c)} = \mathbf{H}_t^{(c)\top}\mathbf{Z}_t^{(c)}$. To handle class reoccurrence in GCIL, $\mathbf{A}^{(c)}$ and $\mathbf{C}^{(c)}$ are maintained per class and updated via recursive least squares.

\begin{algorithm}[t]
\caption{ADIR for GCIL in SSL}
\label{alg:air}
\begin{algorithmic}[1]
\Require Task data $\mathcal{D}_t = \{(\mathbf{x}_i^{(t)}, \mathbf{y}_i^{(t)}, \mathbf{z}_i^{(t)})\}_{i=1}^{N_t}$, frozen feature extractor $F_{\text{MLP}}$, regularization parameters $\gamma_0, \alpha$
\Ensure Weight matrix $\mathbf{{W}}^*_t \in \mathbb{R}^{1000 \times 360}$
\State Initialize $\mathbf{A}^{(y)} \leftarrow \mathbf{0}^{1000 \times 1000}$, $\mathbf{C}^{(y)} \leftarrow \mathbf{0}^{1000 \times 360}$, $N(y) \leftarrow 0$ for all classes $y \in \{1, \dots, 360\}$
\For{each $(\mathbf{x}_i, \mathbf{y}_i, \mathbf{z}_i) \in \mathcal{D}_t$}
    \State $\mathbf{h}_i \leftarrow F_{\text{MLP}}(\mathbf{x}_i) \in \mathbb{R}^{1000}$ \Comment{Extract features}
    \State $y \leftarrow \arg\max(\mathbf{y}_i)$ \Comment{Get class index}
    \State $\mathbf{A}^{(y)} \leftarrow \mathbf{A}^{(y)} + \mathbf{h}_i \mathbf{h}_i^\top$ \Comment{Update auto-correlation}
    \State $\mathbf{C}^{(y)} \leftarrow \mathbf{C}^{(y)} + \mathbf{h}_i \mathbf{z}_i^\top$ \Comment{Update cross-correlation}
    \State $N(y) \leftarrow N(y) + 1$ \Comment{Update class count}
\EndFor
\State Compute $\pi_y \leftarrow 1 / N(y)$ for all $y$ with $N(y) > 0$
\State Compute $\text{Gini}_t$ using Eq.~\eqref{eq:gini} 
\State Set $\gamma_t \leftarrow \gamma_0 \cdot e^{\alpha \cdot (\text{Gini}_t - 0.5)}$ \Comment{Adaptive regularization}
\State Compute $\mathbf{A}_{1:t} \leftarrow \sum_{y=1}^{360} \pi_y \mathbf{A}^{(y)}$, $\mathbf{C}_{1:t} \leftarrow \sum_{y=1}^{360} \pi_y \mathbf{C}^{(y)}$
\State \Return $\mathbf{{W}}^*_t \leftarrow (\mathbf{A}_{1:t} + \gamma_t \mathbf{I})^{-1} \mathbf{C}_{1:t}$
\end{algorithmic}
\end{algorithm}
\vspace{-2mm}

\section{Experiment}
\label{sec:experiment}


\subsection{Dataset}
\label{ssec:dataset}

We use the SSLR benchmark~\cite{he2018deep} where 4-channel audio recordings are sampled at 48 kHz with spatial annotations (all possible microphone pairs are used thus $P$=6). Following the GCIL setting~\cite{air}, the dataset is partitioned into 10 sequential tasks, denoted as $\mathcal{D}_i$, where $i=1,\dots,10$. Each task comprises up to 60 disjoint DoA classes at $1^\circ$ resolution. Task 1 introduces 60 new classes, while tasks 2–9 add 30 new classes and 30 previously encountered classes from earlier tasks to mimic persistent sources~\cite{acil}; task 10 covers the remaining classes.

For input, we extract GCC-PHAT features $\mathbf{x}_i^{(t)} \in \mathbb{R}^{6 \times 51}$ from 170 ms segments, corresponding to 51 delay coefficients per microphone pair. One-hot DoA labels $\mathbf{y}_i^{(t)}$ are used for evaluation, while Gaussian-smoothed targets $\mathbf{z}_i^{(t)}$ ($\sigma=5^\circ$) are adopted for training. Test sets $\mathcal{D}_t^{\text{test}}$ include only the classes of task $t$, and evaluation is performed cumulatively on $\mathcal{D}_1^{\text{test}}, \dots, \mathcal{D}_t^{\text{test}}$ to track forgetting.
\begin{figure}[!b]
\vspace{-1.5em}
\centering
\begin{subfigure}{\columnwidth}
  \centering
  \includegraphics[width=.92\linewidth]{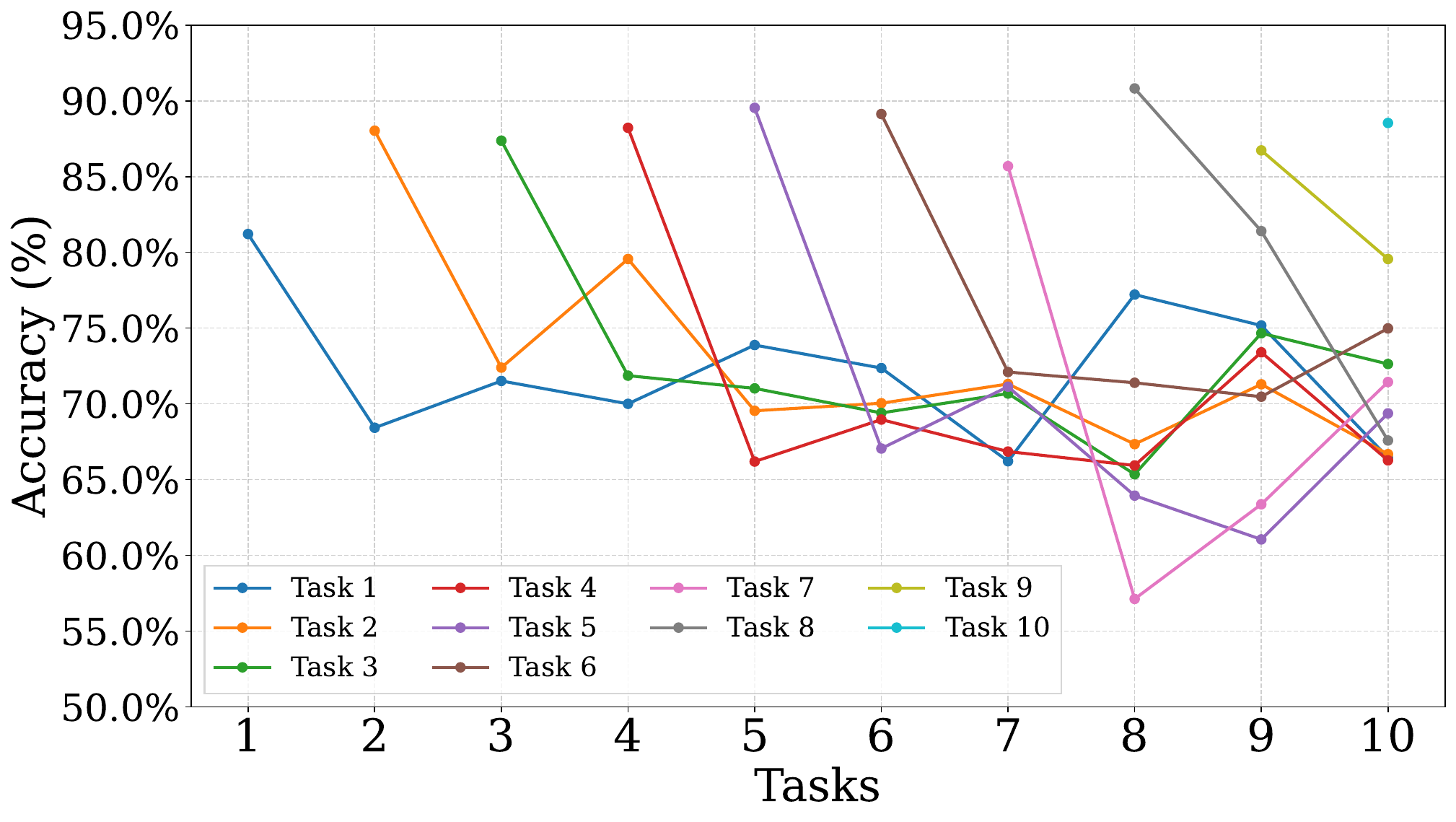}
  \caption*{~(a) Simple MLP (Accuracy scale: 50\%--95\%)}
  \label{fig:lower_bound}
\end{subfigure}

\begin{subfigure}{\columnwidth}
  \centering
  \includegraphics[width=.92\linewidth]{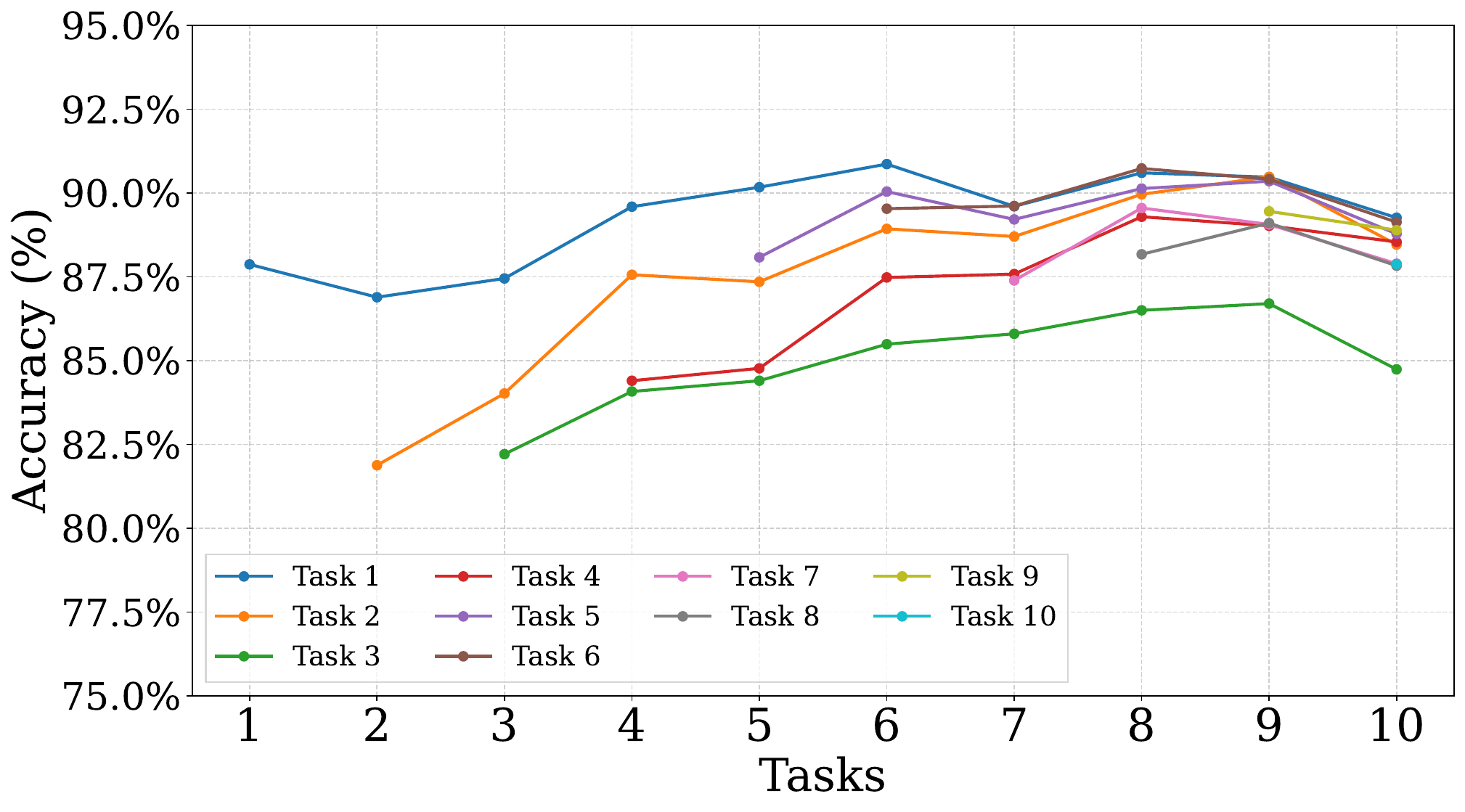}
  \caption*{~(b) Our proposed SSL-GCIL (Accuracy scale: 75\%--95\%)}
  \label{fig:arm_aug}
\end{subfigure}

\caption{Comparison of performance curves between the baseline method and our proposed SSL-GCIL approach. 
}
\label{fig:curve}
\end{figure}
\begin{table*}[t]
\centering
    \def\arraystretch{1.12}
    \setlength{\tabcolsep}{1.33pt}
    \setlength{\abovetopsep}{0pt}
    \setlength\belowbottomsep{0pt} 
    \setlength\aboverulesep{0pt} 
    \setlength\belowrulesep{0pt}

\caption{Performance comparison of different methods under varying SNR conditions. Models are trained on clean data and tested on noisy data. {(`$\uparrow$' indicates that higher scores are preferable while `$\downarrow$' indicates that lower scores are preferable. $^\star$ indicates the upper bound.)}
}
\label{tab:snr_results}
\small
\begin{tabular}{@{}cccccccccccccccccccc@{}}
\toprule
\multirow{2}{*}{SNR} & 
\multicolumn{3}{c}{Joint Training$^\star$} & 
\multicolumn{3}{c}{iCaRL~\cite{icarl}} & 
\multicolumn{3}{c}{LwF~\cite{lwf}} & 
\multicolumn{3}{c}{ACIL~\cite{acil}} & 
\multicolumn{3}{c}{GACL~\cite{gacl}} & 
\multicolumn{3}{c}{Ours} \\
\cmidrule(r){2-4} \cmidrule(r){5-7} \cmidrule(r){8-10} \cmidrule(r){11-13} \cmidrule(r){14-16} \cmidrule(r){17-19}
  & {MAE$\downarrow$} & {ACC$\uparrow$} & {BWT$\uparrow$} & {MAE$\downarrow$}  & {ACC$\uparrow$} & {BWT$\uparrow$} & {MAE$\downarrow$}  & {ACC$\uparrow$} & {BWT$\uparrow$} & {MAE$\downarrow$}  & {ACC$\uparrow$} & {BWT$\uparrow$} & {MAE$\downarrow$}  & {ACC$\uparrow$} & {BWT$\uparrow$} & {MAE$\downarrow$}  & {ACC$\uparrow$} & {BWT$\uparrow$} \\
 
\midrule
\multicolumn{1}{c}{Clean}   & 4.3 & 93.5 & \multicolumn{1}{c}{--} & 5.9 & 85.8 & -3.8 & 6.4 & 76.7 & -12.8 & 6.5 & 85.8 & -1.7 & 6.9 & 78.3 & -1.0 & \textbf{5.3} & \textbf{89.0} &  \textbf{1.6} \\
\multicolumn{1}{c}{20 dB}   & 6.0 & 88.4 & \multicolumn{1}{c}{--} & 7.9 & 78.6 & -6.9 & 9.2 & 69.1 & -13.7 & 8.2 & 81.7 & -1.2 & 9.3 & 71.5 & -1.2 & \textbf{7.3} & \textbf{84.9} &  \textbf{1.0} \\
\multicolumn{1}{c}{10 dB}   & 13.7 & 73.2 & \multicolumn{1}{c}{--} & 27.5 & 46.1 & -7.5 & 20.6 & 61.4 & -11.1 & 15.8 & 64.4 & -1.3 & 17.8 & 64.9 & -1.4 & \textbf{15.5} & \textbf{69.0} &  \textbf{0.9} \\
\multicolumn{1}{c}{0 dB}    & 30.8 & 42.8 & \multicolumn{1}{c}{--} & 38.7 & 33.4 & -4.0 & 42.8 & 28.4 & -5.1 & 38.4 & 35.3 & -1.4 & 38.5 & 33.6 & -1.1 & \textbf{36.9} & \textbf{37.9} &  \textbf{0.8} \\
\multicolumn{1}{c}{-10 dB}  & 47.5 & 18.6 & \multicolumn{1}{c}{--} & 54.9 & 9.2 & -1.5 & 58.6 & 6.2 & -0.4 & \textbf{50.1} & 9.9 & -0.3 & 58.9 & 7.9 & -1.4 & 51.7 & \textbf{16.0} &  \textbf{0.6} \\
\midrule
\multicolumn{1}{c}{Avg.}     & 20.5 & 63.3 & \multicolumn{1}{c}{--} & 27.0 & 50.6 & -4.7 & 27.5 & 48.4 & -8.6 & 23.8 & 55.4 & -1.2 & 26.3 & 51.2 & -1.2 & \textbf{23.3} & \textbf{59.4} &  \textbf{1.0} \\
\bottomrule
\end{tabular}
\end{table*}


To simulate realistic long-tailed dynamics, each task follows an exponential decay distribution with task-specific imbalance. Specifically, the decay rate increases from $\lambda_1=0.05$ to $\lambda_{10}=0.5$, with class counts given by $N_c^{(t)} = \lfloor N_{\text{max}} e^{-\lambda_t (c-1)} \rfloor$, $N_{\text{max}}=500$. This design yields
stronger imbalance across tasks, with varying Gini coefficients to capture long-tailed shifts in real-world SSL.


\vspace{-3mm}
\subsection{Evaluation Metrics}
\label{ssec:metrics}

We evaluate performance using three metrics: {Mean Absolute Error (MAE)}, {Accuracy (ACC)}, and {Backward Transfer (BWT)}. Specifically, \textbf{MAE} measures the average angular error between predicted and true directions of arrival. \textbf{ACC} is the proportion of predictions within a $5^\circ$ tolerance of the true direction following~\cite{chen2024locselect}. \textbf{BWT} quantifies forgetting in incremental learning:
$ \text{BWT} = \frac{1}{T-1} \sum_{k=1}^{T-1} (A_{T,k} - A_{k,k})$
where $A_{k,k}$ is the ACC on task $k$ immediately after training, and $A_{T,k}$ is the ACC on task $k$ after training all $T$ tasks. Positive and negative BWT indicates knowledge retention and forgetting, respectively.

\vspace{-3mm}
\subsection{Baselines}
\label{ssec:baselines}

We compare our method against four SoTA CIL baselines, \textcolor{black}{i.e.,}  \textbf{LwF}~\cite{lwf}, \textbf{iCaRL}~\cite{icarl}, \textbf{ACIL}~\cite{acil} and \textbf{GACL}~\cite{gacl}. We adapted them to the same SSL-GCIL setting as ours. In addition, we report a \textit{lower bound} using a simple MLP trained only on the initial task without any incremental adaptation~\cite{qian2021doa}, and an \textit{upper bound} using joint training on all tasks with access to the full dataset. To ensure fairness, all baselines employ the same MLP feature extractor pretrained on the first task and frozen thereafter. For Task 1, all parameters are optimized with Adam with a learning rate of 0.001, using a binary cross-entropy loss. For subsequent tasks, the feature extractor $\mathbf{W}_{\text{MLP}}$ is kept fixed, and only the classifier weights $\mathbf{W}_{\text{t}}$ are updated under the same training protocol.



\begin{table}[t]
\centering
\caption{Ablation studies under clean conditions (GDA: GCC-PHAT-based data augmentation, ADIR: Adaptive dynamic imbalance rectifier).}
\label{tab:ablation}
\begin{tabular}{@{}ccccc@{}}
\toprule
{GDA} & {ADIR} & {MAE$\downarrow$} & {ACC$\uparrow$} & {BWT$\uparrow$} \\
\midrule
$\text{\texttimes}$ & $\text{\texttimes}$ & 7.5 & 72.0 & -17.7 \\
$\checkmark$ & $\text{\texttimes}$ & 7.4 & 75.0 & -15.8 \\
$\text{\texttimes}$ & $\checkmark$ & 6.1 & 82.4 & 1.4 \\
$\checkmark$ & $\checkmark$ & \textbf{5.3} & \textbf{89.0} & \textbf{1.6} \\
\bottomrule
\vspace{-3.0em}
\end{tabular}
\end{table}

\vspace{-3mm}
\subsection{Results}
\label{ssec:results}

Table~\ref{tab:snr_results} presents the {comprehensive} evaluation results across varying SNR conditions. Under clean conditions, our method achieves the best performance with an MAE of \textbf{5.3$^{\circ}$}, ACC of \textbf{89.0\%}, and positive BWT of \textbf{1.6}, {thereby} indicating effective knowledge retention without catastrophic forgetting. Compared to the strongest baseline ACIL, we achieve \textbf{1.2$^{\circ}$} lower MAE, \textbf{3.1\%} higher ACC, and \textbf{3.3} higher BWT. The joint training upper bound (MAE: 4.3$^{\circ}$, ACC: 93.5\%) represents the performance ceiling when all data is available simultaneously, while our method approaches this ideal scenario with only incremental access.

As noise levels increase, all methods experience performance degradation, while our approach consistently maintains the relative advantage. For example, at 20dB SNR, we achieve 7.3$^{\circ}$ MAE and 84.9\% ACC, {significantly} outperforming ACIL by 0.9$^{\circ}$ and 3.2\% {respectively}. Under severe -10dB conditions, our method achieves the highest ACC (16.0\%) among other incremental learning approaches, demonstrates remarkable noise robustness.

The BWT metric reveals our method's {exceptional} resistance to catastrophic forgetting. While {all} baselines show negative BWT, our method maintains positive BWT across all noise levels, with 0.6 even at -10dB SNR. This confirms that our adaptive regularization and augmentation strategy effectively preserves previously learned knowledge {in challenging environments}.

Fig.~\ref{fig:curve}(a) illustrates the severe catastrophic forgetting problem of the baseline method~\cite{qian2021doa}, where the accuracy on previously learned tasks drops significantly as new tasks are introduced. For instance, after training on Task 10, Task 1's accuracy drops to 66.5\%, showing a decline of nearly 15\% from its initial performance. In contrast, our method (Fig.~\ref{fig:curve}(b)) maintains stable performance across all tasks, with accuracies consistently above 84\% even after incremental training. The ablation studies in Table~\ref{tab:ablation} confirm that both components contribute significantly: ADIR alone improves BWT from -17.7 to 1.4, effectively mitigating negative backward transfer, while adding GDA further boosts ACC from 82.4\% to 89.0\%. The combination achieves the best balance between accuracy and stability. Overall, our proposed method consistently outperforms all baselines in most scenarios, demonstrating superior robustness to incremental learning challenges and environmental noise.

\vspace{-2mm}
\section{Conclusion}
\label{sec:conclusion}


In this paper, we present a novel SSL framework under GCIL that integrates GDA with ADIR, thus successfully alleviates class imbalance in long-tailed distributions and mitigates catastrophic forgetting without requiring exemplar storage. Specifically, the GDA module effectively addresses intra-task imbalance by synthesizing tail-class samples through peak statistics manipulation, while the ADIR component mitigates inter-task imbalance via task-adaptive regularization guided by Gini coefficients. On the SSLR benchmark, our model achieves state-of-the-art results, with 89.0\% accuracy, a mean absolute error of 5.3$^\circ$, and a backward transfer of 1.6, consistently outperforming all baselines. 
For future work, we plan to explore extensions of integrating audio-visual signals for SSL and the online adaptation strategy.
\bibliographystyle{IEEEbib}
\bibliography{strings,refs}
\end{document}